\documentclass[epjST]{svjour}
\usepackage{graphics}
\begin{document}
\title{Dynamical nonextensivity or nonextensive dynamics?}
\author{Jacek Ro\.zynek\thanks{\email{jacek.rożynek@ncbj.gov.pl}} \and  Grzegorz Wilk\thanks{\email{grzegorz.wilk@ncbj.gov.pl}}}
\institute{National Centre for Nuclear Research\\
 Department of Fundamental Research, 02-093 Warsaw, Poland}
\abstract{
Dense matter is usually described using some kind of mean field theory (MFT) model based on Boltzmann-Gibbs (BG) extensive statistics. However, in many cases the conditions justifying the use of BG statistics are not fulfilled because the systems considered are explicitly nonextensive. In such cases one either enriches the original MFT by adding some dynamical elements violating extensivity (like, for example, long range correlations or intrinsic fluctuations), or one replaces the BG statistics by its nonextensive counterpart characterized by some nonextensivity parameter $q$ ($q\neq 1$ and for $q \to 1$ one returns to the extensive situation). In this work, using a simple quasi-particle description of dense matter (with interaction modelled by effective fugacities, $z$) we discuss the mutual interplay of non-extensiveness and dynamics (i.e., $q$ and $z$).
}
\maketitle
\section{Introduction}
\label{intro}

Dense matter is customarily described using some variation of the mean field theory approach (MFT) based on the extensive Boltzmann-Gibbs statistics (BS) (see \cite{Symmetry} for a recent review and references). However, in most cases the systems considered are not extensive because there are phenomena like, for example, long-range correlations or intrinsic fluctuations, not accounted for by the MFT used; in such cases the conditions justifying the use of BG statistics are not fulfilled. The usual remedy in such cases is to enrich the original dynamics by adding to the initial MFT some new elements accounting for these factors. The other possibility is to keep the original MFT intact, but to replace the BG statistics by its nonextensive counterpart which, by definition has these factors built in; usually by the Tsallis statistics (TS) \cite{PSA,PSA1,CSRH,JRGW1,JRGW2,JRGW3}. It is characterized by a nonextensivity parameter $q\neq 1$ such that, for $q \to 1$ one returns to the extensive situation of BG statistics. In such an approach modifications caused by introduction of nonextensive statistics are supposed to sum up the actions of all factors violating extensivity, both dynamical and caused by the environment. It is therefore expected that when these factors are gradually identified and their impact is accounted for by a suitable modification of the dynamics of the original model, the $q$ needed to fit data gradually tends to unity and one recovers the usual extensive situation of the BS, albeit now used with a modified MFT \cite{MB}. Note that this means that there is no such thing as a non-extensive free particle, and this is an inherent dynamical feature of  nonextensivity \cite{Biro-1,Biro-2,Biro-3}. To some extent the nonextensive statistics replaces the nonextensive dynamics (and vice versa), therefore in many cases we can talk interchangeably about {\it dynamical nonextensivity} or {\it nonextensive dynamics}. In this paper we shall take a closer look at this feature. To facilitate this task we limit ourselves to the simplest possible implementation of the dynamics in the form of some specific quasi-particle model of interactions proposed in \cite{QPM,QPMu} in which interactions are modelled by only one (albeit temperature dependent) parameter for each type of particle (here quarks, $q$, and gluons, $g$), and the effective fugacities,$z_{i=q,g}$, cf. \cite{Symmetry} for more references). In such an approach a single nonextensivity parameter, $q$, can be confronted with a set of individual dynamical parameters, $z_{i=q,g}$.

\section{The quasi-particle model in extensive and nonextensive environments}
\label{sec:form-qz}

For completeness of presentation we start with a short reminder of the essentials of the quasi-particle model in extensive, $z$-QPM, and non-extensive, $qz$-QPM, environments (cf. \cite{Symmetry} for details and further references).

In the case of an extensive environment the $z$-QPM \cite{QPM} is based on the effective equilibrium distribution function for quasi-partons ($i=q,~s,~g$ for, respectively, $u$ and $d$ massless quarks, strange quarks of mass $m$ and gluons; $e(x) = \exp(x)$, $\xi = +1$ for bosons and $-1$ for fermions and $\beta = 1/T$):
\begin{eqnarray}
n\left[ x^{(i)}\right] &=& \frac{1}{\frac{1}{z^{(i)}}e\left[ x^{(i)} \right] - \xi} = \frac{1}{e\left[ \tilde{x}^{(i)}\right] - \xi}, \label{fex-d}\\
x^{(i)} &=& \left\{
 \begin{array}{lll}
  \beta \left[ E_i - \mu^{(i)}\right]~~ &~~ {\rm if}&~~ i=q, s,\\
  \beta E_i ~~&~~ {\rm if} &~~ i= g.
  \end{array}\right.\qquad {\rm and}\qquad \tilde{x}^{(i)} = x^{(i)} - \ln z^{(i)}(\tau),
  \label{x(i)}
\end{eqnarray}
$E_{i=q,g}=p$ and $E_s =\sqrt{m^2 + p^2}$. Note that $e(x)\cdot e(-x) =1$, a consequence of which is that $n(x) + n(-x) = \xi$. The $z^{(i)} \leq 1$ denote the effective fugacities which describe the interactions and, by assumption, depend only on the scaled temperature, $\tau = T/T_{c}$ ($T_{c}$ is the temperature of transition to the deconfined phase of QCD). They are obtained by fits to the lattice QCD data, for  $z^{(i)} = 1$ one has free particles. In its original version $z$-QPM is formulated for zero chemical potential $\mu^{(i)}$ \cite{QPM} (reflecting difficulties with accounting for it in lattice QCD calculations; nonzero $\mu$, not connected with $z$, was introduced recently in \cite{QPMu} anticipating expected future developments in lattice QCD calculations).

In the case of $qz$-QPM one takes some {\it extensive system of quasi-particles}, the interactions of which are described by fugacities $z^{(i)}$ (and, possibly, also by some chemical potentials $\mu^{(i)}$, not connected with these fugacities), and immerses it in a nonextensive environment characterised by a nonextensivity parameter $q\neq 1$. Note that, as mentioned before, this means that we are dealing now with particles which are not really free, even when the dynamics is switched off. As a result one gets a {\it nonextensive system of interacting quasi-particles} and, assuming that the external dynamical information encoded in the results of the lattice QCD simulations remains intact, one has to find a new set of fugacities, $z^{(i)}_q$, which, together with modifications in the distribution caused by $q\neq 1$, will reproduce this information. Technically speaking, one simply replaces in Eq. (\ref{fex-d}) the exponential function $e(x) = \exp(x)$ by its nonextensive equivalent, the nonextensive exponent $e_q(x) = \exp_q(x)$, and its dual, $e_{2-q}(-x) = \exp_{2-q}(-x)$, defined as:
\begin{equation}
e_q(x) = [ 1 + (q-1)x ]^{\frac{1}{q-1}},\qquad e_{2-q}(-x) = [1 + (1-q)(-x)]^{\frac{1}{1-q}}, \label{eqe}
\end{equation}
$e_q(x) \to e(x)$ and $e_{2-q}(-x) \to e(-x)$ for $q\to 1$, the correspondig $q$ and $(2-q)$-logarithm functions are:
\begin{equation}
\ln_q X = \frac{X^{q-1} - 1}{q - 1} \stackrel{q \rightarrow 1}{\Longrightarrow} \ln X\quad {\rm and}\quad \ln_{2-q}X = \frac{X^{1-q}-1}{1-q} \stackrel{q \rightarrow 1}{\Longrightarrow} \ln X.  \label{lnq}
\end{equation}
As a result one obtains the following nonextensive particle occupation numbers:
\begin{equation}
n_q\left[ \tilde{x}^{(i)} \right] = \frac{1}{ e_q\left[ \tilde{x}^{(i)}\right] - \xi} = \frac{e_{2-q}\left[- \tilde{x}^{(i)}\right]}{1 - \xi e_{2-q}\left[- \tilde{x}^{(i)}\right]}, \label{enq}
\end{equation}
with $\tilde{x}^{(i)}$ given by Eq. (\ref{x(i)}) (with energy, $E_i$, and chemical potential, $\mu$, remaining unchanged). Note that now  $e_q(x)\cdot e_{2-q} =1$ and, respectively, $n_q(x) + n_{2-q}(-x) = \xi$. To preserve thermodynamic consistency in the nonextensive environment, one has to use effective occupation numbers in the form of $\left[ n_q(x) \right]^q$ \cite{Symmetry}. In all calculations one has always to ensure that the corresponding $q$-exponents are nonnegative and real valued, cf. \cite{Symmetry} for details.

\section{The interplay between the fugacity $z$ and the nonextensivity $q$ - dynamics vs nonextensivity}
\label{sec:Interplay}

In such a QPM the investigation of the interplay between the dynamics and nonextensivity comes to the study of the interplay between the fugacities $z_{i=q,g}$, representing the dynamics, and the nonextensivity $q$, representing the action of the environment (always keeping in mind that $q\neq 1$ combines the action of all factors causing nonextensivity, even in the absence of any interaction, i.e., when $z_{i=q,g} = 1$, and that the nonextensive particles cannot be considered as being completely free \cite{Biro-1,Biro-2}). Note first that both $z$ and $q$ deform the original Fermi-Dirac or Bose-Einstein distributions of noninteracting particles, but they do it in two different, incompatible ways. The $z$-deformation is {\it local}, it is supposed to depend only on the scaled temperature $\tau = T/T_{crit}$. As can be seen in Eq. (\ref{fex-d}) the action of $z = z\left( \tau=T/T_{crit}\right)$ is, technically, the same as action of some kind of "artificial chemical potential", $ \mu_z = T\ln z(\tau)$ (which adds to the usual chemical potential). This means that the form of the distribution remains intact, only its argument changes (therefore adding some true chemical potential $\mu$ to the original $z$-QPM can be combined with fugacity $z$ and results in some new "effective" $\tilde{z}$). In contrast to this the $q$-deformation is {\it global} and, in principle, the parameter $q$ is assumed to be independent of temperature\footnote{At least as long as there is no energy exchange between the heat bath and the environment, this would result in the replacement $T \to T=T_{effective} = T(q)$ (so far this has been shown only for the Boltzmann statistics \cite{WW_Tq,WW_Tq1}).}. This means that one cannot fully replace $z$ by $q$ (and vice versa). To see it better let us identify the effective extensive (i.e., calculated for $q=1$) occupation number considered as a function of fugacity $z$, with its nonextensive counterpart defined mainly by a nonextensivity parameter $q$ (with no other dynamical effects, i.e., with $z=1$, note that it is given not by $n_q$, but by $n^q_q$):
\begin{equation}
n(x;z,q=1) = \frac{1}{\frac{1}{z}e(x) - \xi} = \left[ n_q(x;z=1,q)\right]^q =\left[ \frac{1}{e_q(x) - \xi}\right]^q.   \label{equality}
\end{equation}
The immediate result is the relation between $z$ and $q$,
\begin{equation}
z = z(q) = \frac{ e(x) }{ \left[ e_q(x) - \xi\right]^q + \xi} \stackrel{q \to 1}{\Longrightarrow} 1.\label{z-funq}
\end{equation}
The reverse relation, $q=q(z)$, is a transcedental function, not accessible analytically. However, relation (\ref{equality}) would imply that the resulting fugacity must depend not only on the nonextensivity $q$, but also on the energy, chemical potential and temperature, i.e., on $x$, $z = z(q;E,\mu,T)$. This is unacceptable because fugacity was introduced to model the dynamics which cannot depend on $x$.

The same argument also precludes the seemingly more general formulation of the $qz$-QPM discussed in \cite{JRGW3}. One could, for example, start with a system of  particles the interactions of which are described by some chemical potentials $\mu^{(i)}$ (not connected with any fugacities, $z=1$) and which are immersed in some nonextensive environment characterized by a nonextensive parameter $q\neq 1$. Adding now to such a system some additional interaction by means of fugacities $z^{(i)}_q$, results in
\begin{eqnarray}
n_q\left[ x^{(i)} \right] &=& \frac{1}{ \frac{1}{z_q^{(i)}}e_q\left[ x^{(i)}\right] - \xi} = \frac{1}{e_q\left[ x_q^{(i)}\right] - \xi}, \label{zenq}\\
x_q^{(i)} &=& \beta \left[ E^{(i)}_q - \tilde{\xi}\mu_q^{(i)}\right]\! -\! \zeta_q^{(i)} =
x^{(i)} \left[ z_q^{(i)}\right]^{1-q}\! -\! \zeta_q^{(i)} = \ln_q\left[ \frac{e_q(x)}{z_q}\right], \label{xenq}
\end{eqnarray}
where $E_q^{(i)} = E_i\cdot\left[ z_q^{(i)}\right]^{1-q}$, $\mu_q^{(i)} = \mu^{(i)}\cdot\left[ z_q^{(i)}\right]^{1-q}$ and $\zeta_q^{(i)} = \ln_{2-q}\left[z_q^{(i)}\right]$, i.e., the energy and chemical potentials become $q$-dependent quantities. However, such induced $q$-dependence of the initial energy and chemical potential is rather unphysical and precludes further applications of this approach.

\begin{figure}[b]
\hspace{-3mm}
\resizebox{0.58\textwidth}{!}{%
  \includegraphics{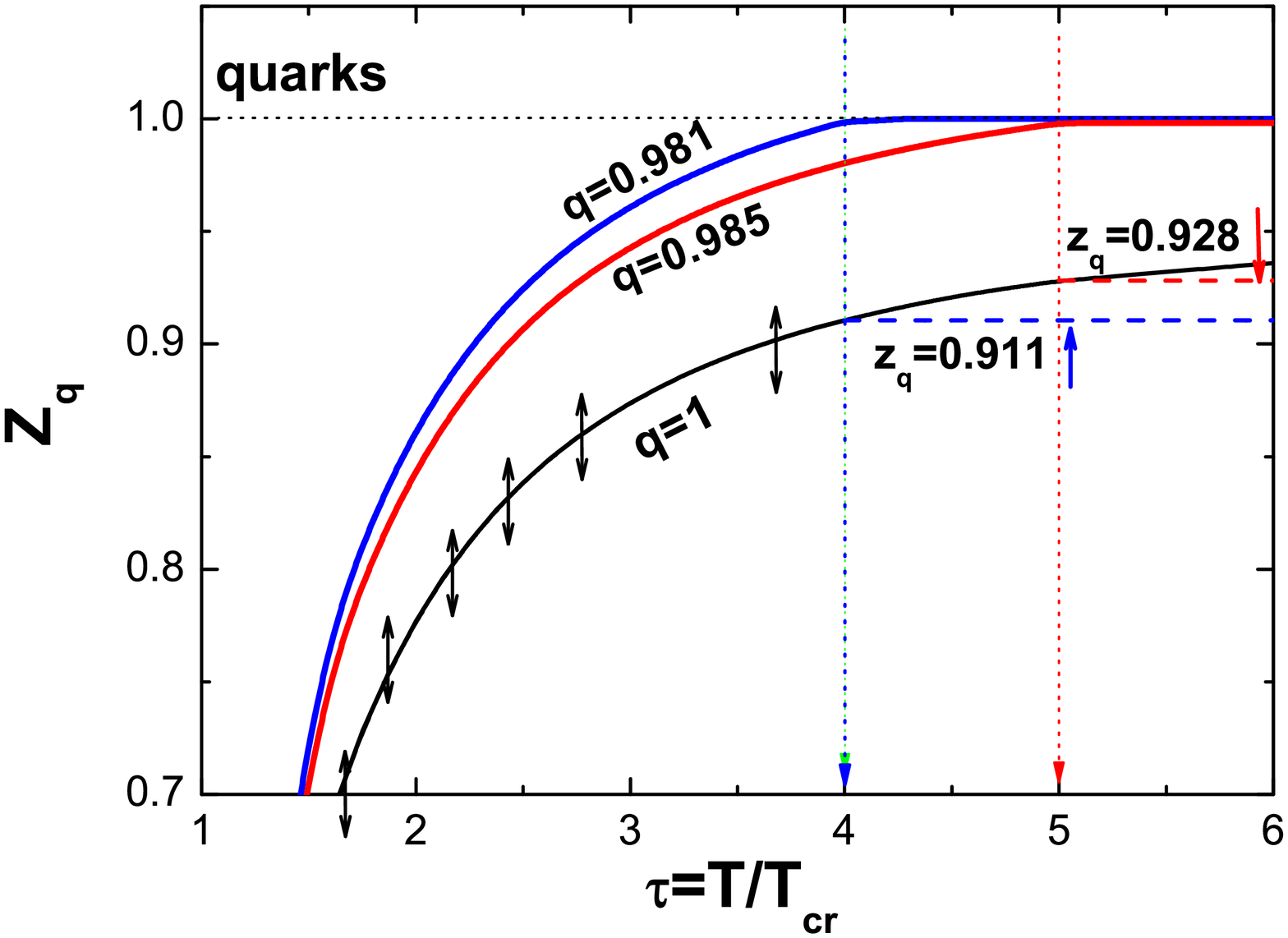}
}
\hspace{-12mm}
\resizebox{0.58\textwidth}{!}{%
  \includegraphics{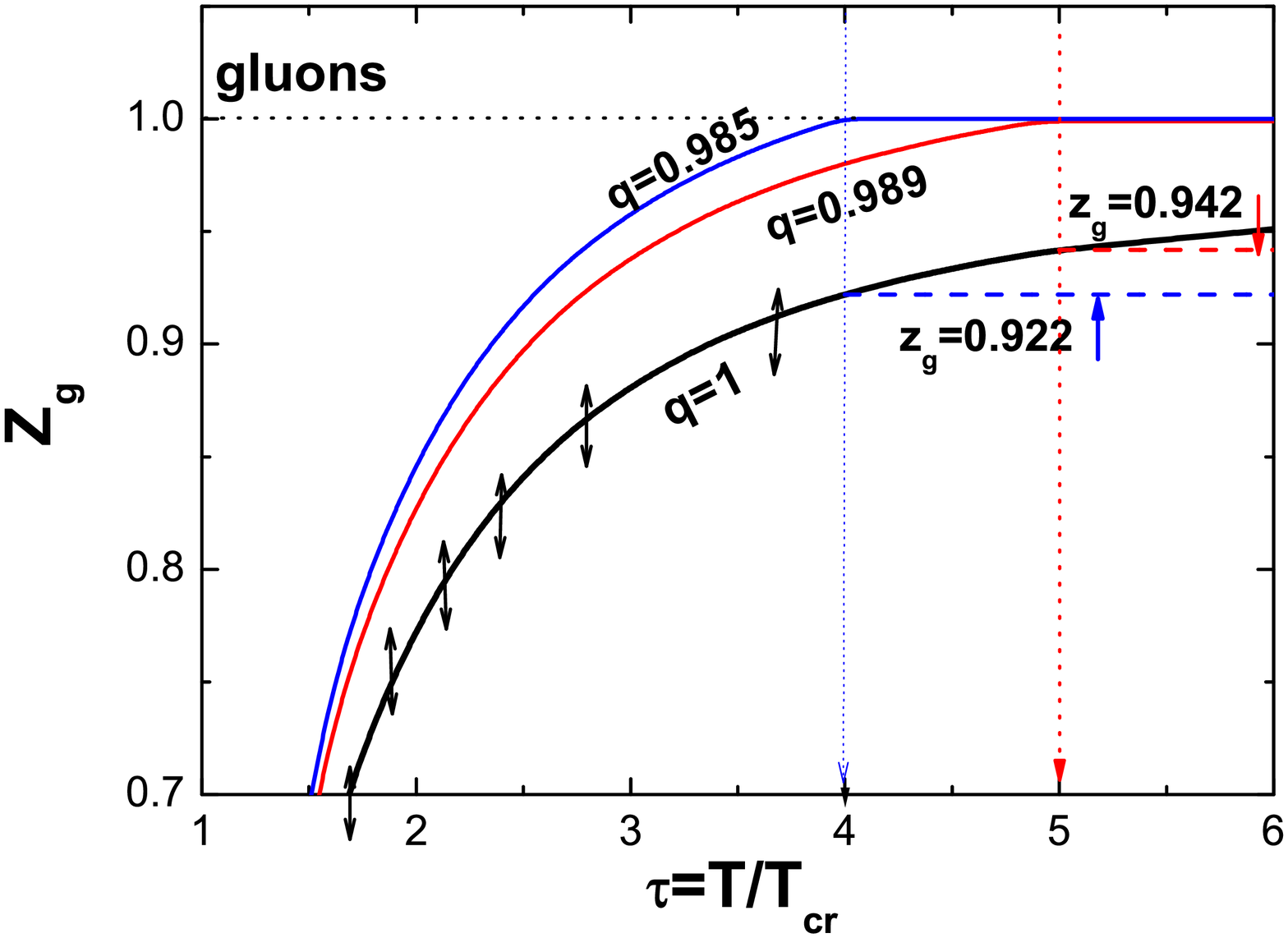}
}
\caption{(Color online) Fugacities $z_{(q,g})$ (left and right panels, respectively) obtained in the $z$-QPM \cite{QPM} (with $q=1$, black curves; the lattice QCD results which were fitted are also shown by arrows \cite{QPM}) as functions of the relative temperature $\tau$. They are compared with the nonextensive fugacities $z_q$ (red and blue curves) obtained from Eqs. (\ref{zq}) and (\ref{F}) for, respectively, $q=0.985$ and $0.981$ for quarks and for $q=0.989$ and $0.985$ for gluons. They were chosen in such a way as to get $z_{(q,g)} = 1$ for, respectively, $\tau = 4$ and $\tau=5$, which are also the values of $\tau$ onwards from which we assume that $z(\tau)$ remains constant.}
\label{Rzqg}
\end{figure}

In \cite{Symmetry,JRGW3} we have discussed in detail how the already known dynamics, represented by effective fugacities $z_{i=q,g}$ introduced in \cite{QPM}, is modified by immersing the system under consideration in some nonextensive environment characterized by the parameter $q\neq 1$. Fugacities $z_{i=q,g}$ were obtained from the results of lattice QCD simulations (which were treated as input data) by comparing the pressures of the gluons and quarks (expressed as functions of the fugacities) with the corresponding pressures obtained from the lattice calculations \cite{QPM}. As a result the effective fugacities were received as functions of scaled temperature, $\tau = T/T_{cr}$, with $T_{cr}$ being the critical temperature. To describe the lattice QCD data over the whole range of $\tau$ considered (i.e., for $\tau < 4$), the $\tau$ range had to be divided in two sectors, each of which was parameterized by a different functional form with the cross-over point at $\tau_g = 1.68$ for gluons and $\tau_q=1.7$ for quarks:
\begin{equation}
z^{(q,g)}(\tau) = a_{(q,g)}\exp\left[ - \frac{b_{(q,g)}}{\tau^5}\right]\Theta\left( \tau_{(q,g)} - \tau\right) + a'_{(q,g)}\exp\left[ - \frac{b'_{(q,g)}}{\tau^2}\right]\Theta\left( \tau - \tau_{(q,g)}\right).\label{zzz}
\end{equation}
The fit parameters are: $a_{(q,g)} = (0.810,~0.803)$, $a'_{(q,g)} = (0.960,~0.978)$, $b_{(q,g)} = (1.721,~1.837)$, $b'_{(q,g)} = (0.846,~0.942)$. The resulting $z^{(q,g)}(\tau)$ are shown in Fig. \ref{Rzqg}) (black curve) where we have also shown the positions of the points obtained from lattice QCD (by arrows) \cite{QPM}. Note that $z^{(q,g)}(\tau \to \infty) = a'_{(q,g)} < 1$. This result, when treated seriously, indicates that with increasing temperature $\tau$ the system of quarks and gluons considered in the QCD lattice simulations never becomes a gas of free streaming non-interacting particles. From the nonextensive point of view, presented below, this would mean that lattice QCD keeps a memory of the interaction and describes a quark-gluon system which is intrinsically nonextensive, with $q<1$, as will be shown below.

The same pressures of the gluons and quarks obtained from the lattice QCD results must now be reproduced in the nonextensive circumstances defined by the nonextensivity parameter $q$, i.e., by using nonextensive particle occupation numbers $n_q\left[ \tilde{x}^{(i)}\right]$ given by Eq. (\ref{enq}) with nonextensive effective fugacities $z_{q\neq 1}$. As in \cite{Symmetry}, we used for this purpose the same parameterization of $z_q(\tau)$ as was used for $z(\tau)$ in Eq. (\ref{zzz}), but with the $q$-dependent values of parameters $(a,b)$ and $(a',b')$ obtained from the same lattice QCD results. The results turned out to be very sensitive to the amount of nonextensivity imposed, limiting our considerations to $|q-1| <<1$. Because the corresponding changes in fugacities, $\delta z_q = z_{q\neq 1} - z_{q=1}$, for these values of $q$ were also small, $\delta z_q <1 $, the exact formulas practically coincided with a linear in $(q-1)$ approximation \cite{Symmetry} (for clarity of presentation we suppress indices $i=q,g$):
\begin{eqnarray}
z_q &\simeq& z_{q=1} + \delta z_q = z_{q=1} \left[ 1 + (1-q) \cdot F\left(q=1,z_{q=1}\right)\right], \label{zq}\\
&&F = \frac{\int_0^{\infty} dp p^2 \left\{ \ln^2[ 1 - \xi e\left( -x;z_{q=1}\right)] + n\left(x;z_{q=1}\right) x^2\right\}}{2\int_0^{\infty} dp p^2 n\left(x;z_{q=1}\right)}. \label{F}
\end{eqnarray}
As $F > 0$, this means that we always have $z_{(q>1)} < z_{(q=1)} < z_{(q<1)}$.

We shall concentrate now on the $q<1$ case, the example of which is displayed in Fig. \ref{Rzqg}. As already mentioned before, it looks like that for $q=1$ (the extensive case) the fugacity never reaches the value $z=1$ corresponding to a noninteracting gas of quarks and gluons, at least not in the considered range of $\tau < 4$ (which is otherwise quite big), one always has an interacting (confined?) system of quasiparticles.  A closer look reveals that this is because for $z=1$ the pressure in such a gas would exceed the pressure obtained from the lattice calculations. From our experience with a nonextensive version of the Nambu-- Jona-Lasinio model of QCD matter \cite{JRGW3} we know that this pressure is reduced when the system becomes nonextensive with $q<1$. When applying it here, using Eqs. (\ref{zq}) and (\ref{F}), one indeed observes that the respective $z_{(q<1)}$ exceeds $z_{(q=1)}$ for the same values of $\tau$ in such a way as to keep up with the original lattice QCD pressure, and very soon they reach, for a given $\tau$, the value $z_{(q<1)}=1$, and it will exceed it for larger values of $\tau$. This means that our system becomes a system of non-interacting nonextensive quasiparticles (noninteracting in the sense that the only interaction is that provided not by the dynamics but by the nonextensivity $q$ \cite{Biro-1,Biro-2,Biro-3}). Decreasing $q$ further for this $\tau$ (or increasing $\tau$ while keeping the same $q<1$) would result in $z_{(q<1)}$ exceeding unity, which we consider unrealistic (because it would correspond to interactions not present in QCD). Therefore, once the $z_{(q<1)}(\tau)$ reaches unity we assume, in what follows, that it remains unity for larger values of $\tau$ (and for the same $q$, as in Fig. \ref{Rzqg})).

In what concerns the examples shown in Fig. (\ref{Rzqg}), note that the range of $\tau$ considered in the lattice QCD simulations was limited, with the last point located at $\tau= 3.7$ \cite{QPM}. This means that extrapolation of $z(\tau)$ obtained from fits to lattice QCD results, cf. Eq. ({\ref{zzz}), to $\tau >> 3.7$, is highly uncertain. Because of this, we assume that at some $\tau$ the fugacity stops increasing and remains constant thereafter. As an example, we choose for further consideration two choices: $\tau =4$ and $\tau =5$. In other words, we tacitly assume that, starting from these values of $\tau$, the QCD interactions remain essentially constant (or decrease very slowly) therefore the corresponding fugacities remain virtually the same: $z_q = z_q(\tau=4)$ or $z_q = z_q(\tau=5)$, respectively, cf. Fig. (\ref{Rzqg}) (as we will see in a moment, this will allow the determination of the non-extensiveness of our QCD system of quarks and gluons). For each of these two values of $\tau$ the respective nonextensivities  $q$ were found (separately for quarks and gluons) such that the values of the corresponding $z_q(\tau)$ reach unity (corresponding to "free, nonextensive gas"). As was the case of the nonextensive $z(\tau)$, we assume that for $\tau > 4$ (or $\tau >5$)  the corresponding values of $z_q(\tau)$ remain unity. As see in Fig. \ref{Rzqg}) the respective values of $q$ are, for quarks, $q=0.981$ for $\tau = 4$ and $q = 0.985$ for $\tau = 5$ (for gluons they are only slightly larger). We therefore argue that the nonextensivity of quarks and gluons is in the range $0.98 < q < 0.99$.

For the case of $q>1$, not shown in Fig. (\ref{Rzqg}), the corresponding $z_q(q,g)$ are always below the $q=1$ curve for the full range of $\tau$. This is because in this case the pressure is increased, therefore the strength of the dynamical interactions (given by $z_q$) has to be reduced. This is a reflection of the fact that in both cases considered the non-extensiveness comes from different sources which work towards confinement for $q<1$ but work against it for $q>1$ (for example, by introducing some extra intrinsic fluctuations). In other words: because in lattice QCD one always has $z<1$ (which reflects the fact that it describes, in fact, more or less confined systems), in a possible equivalent nonextensive approach this would correspond to a nonextensivity parameter smaller than unity, $q<1$ (which in the colloquial understanding of non-extensiveness corresponds to broadly understood correlations). The $q>1$ type of nonextensivity corresponding (again, broadly speaking) to some kind of effective, intrinsic, fluctuations manifesting themselves as fluctuations of the temperature $T$, would demand $z>1$, which is forbidden in lattice QCD. On the other hand, if taken seriously, one could probably argue that this comes not so much from the QCD dynamics, but rather from the environment (and would, figuratively speaking, come from the heterogeneity of the heat bath and from possible energy transfers to and from it \cite{WW_Tq,WW_Tq1}.

\section{Summary}
\label{sec:SC}

Generally speaking, systems for which the fugacities $z = 1$ in the limit of large $T$ are extensive systems where a single-particle description works without an additional field persisting for high temperatures (without long range interactions). If, for $T\to \infty$, the fugacity $z<1$ our system is nonextensive with reduced pressure, whereas for $z>1$ in this limit it is nonextensive with increased pressure (always in comparison with the situation for the $z=1$ case). This means that we can adjust the system pressure in models with fugacity to determine its model non-extensivity.

Because, as already mentioned, the action of nonextensivity $q$ has a {\it global} character whereas that of the fugacity $z$ is {\it local}, therefore, in principle they are complementary and are not substitutable for each other. To put it differently, usually fugacity $z$ models phenomenologically dynamics of the mean field theory type (MFT) in which there are no correlations and intrinsic fluctuations, whereas these two features are exactly those described by the nonextensivity $q$. Therefore, the results obtained here should be considered as some illustration of how to introduce effects of correlation or fluctuations into Finally,MFT models, or, vice versa, how to add to the system with correlations and/or fluctuations a certain smoothing factor for these effects.

Finally, let us mention a recent work \cite{HJ} in which the authors use numerical field theoretical simulations to calculate particle yields. It turned out that in the model of local particle creation one observes rather natural deviations from the pure exponential distributions towards Tsallis like $q$-exponents. Interestingly, for a quantum SU(3) Yang-Mills gauge theory applied to gluons (which could perhaps be a good replacement for our QCD) they obtained similar values of $q<1$ and similar dependence on the temperature as we observe in our case. The $q>1$ was obtained for a toy model of classical $\Phi^4$ theory. From our point of view  the natural explanation would be the presence of some kind of confinement in the first case (which would correspond to our lattice QCD situation) and its lack in the second case (in our case it would correspond to immersing our system in a heat bath which is so nonuniform that it overcomes the confinement forces).
\vspace{1cm}

The research of GW was supported in part by the National Science Center (NCN) under contract 2016/22/M/ST2/00176. We would like to thank warmly Dr. Nicholas Keeley for reading the manuscript.

\end{document}